\documentclass[10pt,twocolumn,aps,english,pra,superscriptaddress]{revtex4} 
\usepackage{amsfonts}
\usepackage[T1]{fontenc}
\usepackage[latin9]{inputenc}
\usepackage{amsmath}
\usepackage{amssymb}
\usepackage{graphicx}
\usepackage{babel}
\usepackage{booktabs}
\usepackage{comment}
\usepackage{tabu}
\usepackage{subfigure}
\usepackage{physics}
\usepackage{extarrows}
\usepackage{hyperref}
\usepackage{xcolor}

\makeatletter
\@ifundefined{textcolor}{}
{
 \definecolor{BLACK}{gray}{0}
 \definecolor{WHITE}{gray}{1}
 \definecolor{RED}{rgb}{1,0,0}
 \definecolor{GREEN}{rgb}{0,1,0}
 \definecolor{BLUE}{rgb}{0,0,1}
 \definecolor{CYAN}{cmyk}{1,0,0,0}
 \definecolor{MAGENTA}{cmyk}{0,1,0,0}
 \definecolor{YELLOW}{cmyk}{0,0,1,0}
 }
\makeatother

\begin{document}

\title{Superresolution in interferometric imaging of strong thermal sources}

\author{Yunkai Wang} 
\author{Yujie Zhang} 
\author{Virginia O.~Lorenz} 
\email{ vlorenz@illinois.edu}
\affiliation{IQUIST, University of Illinois at Urbana-Champaign, Urbana, IL 61801 USA}
\affiliation{Department of Physics, University of Illinois at Urbana-Champaign, Urbana, IL 61801 USA}

\date{\today }

\begin{abstract}
Imaging using interferometer arrays based on the Van Cittert-Zernike theorem has been widely used in astronomical observation. Recently it was shown that superresolution can be achieved in this system for imaging two weak thermal point sources. Using quantum estimation theory, we consider the fundamental quantum limit of resolving the transverse separation of two strong thermal point sources using interferometer arrays, and show that the resolution is not limited by the longest baseline. We propose measurement techniques using linear beam splitters and photon-number-resolving detection to achieve our bound. Our results demonstrate that superresolution for resolving two thermal point sources of any strength can be achieved in interferometer arrays.
\end{abstract}

\maketitle

\textit{Introduction}- Rayleigh's limit, which says that with a single lens imaging system we cannot resolve sources with separation less than the diffraction-limited spot size of the point spread function, has been widely used to quantify imaging resolution \cite{rayleigh1879xxxi}. Rayleigh's limit was revisited from the perspective of quantum metrology by Tsang \textit{et al.}~\cite{tsang2016quantum,tsang2019resolving}, who showed that Rayleigh's limit for estimating the separation of two weak incoherent point sources using a single lens can be overcome with more carefully designed measurement strategies. From the quantum metrology perspective, Rayleigh's limit is a consequence of an improperly chosen measurement that both restricts access to information and destroys information by collapsing the state. A quantum-mechanical description of imaging allows us to design a measurement to access information about the states through use of prior information about the source, such as the assumption of point sources of equal intensity~\cite{tsang2016quantum}.

Since this ground-breaking result was announced, studies have shown that measurements constructed to avoid Rayleigh's limit for estimating the separation of two weak thermal point sources of equal strength using a finite-sized single lens are not unique. Strategies include the original proposal based on projection onto the Hermite-Gaussian spatial modes \cite{tsang2016quantum} and others based on using an image-inversion interferometer \cite{nair2016interferometric}, adding a phase plate before half of the image \cite{tham2017beating}, using an array of homodyne detectors \cite{datta2020sub}, and exploiting Hong-Ou-Mandel interference using two copies of the incoming photonic state \cite{parniak2018beating}. Experimentally, the quantum theory of superresolution has been verified \cite{zhou2019quantum,tham2017beating}. Toward more realistic applications, there are several practical issues that must be addressed. The constructed optimal measurement can rely on prior information on the source, such as the centroid of the two point sources \cite{tsang2016quantum}. Overhead photons can be used to obtain this prior information, and more carefully designed adaptive measurements can still provide an advantage over conventional imaging \cite{grace2019approaching}. If there exists other noise besides the photon shot noise, the sensitivity of the measurement will be degraded, as discussed in Refs.~\cite{llen2020resolution,lupo2020subwavelength}.

It is of interest to expand the quantum-mechanical description of imaging to cover more types of sources and imaging modalities. Work on other sources in the case of single-lens imaging includes studying point sources of stronger strength \cite{lupo2016ultimate,nair2016far} and unequal strength \cite{vrehavcek2017multiparameter,vrehavcek2018optimal}, estimating point source locations in two and three dimensions \cite{ang2017quantum,yu2018quantum,napoli2019towards,prasad2019quantum}, and finding the sensitivity limit of imaging a more general extended source \cite{dutton2019attaining,zhou2019modern,tsang2019quantum,tsang2019semiparametric,tsang2019quantumsemi,lupo2020quantum}.  
As a first step to study imaging systems other than a single lens, Ref.~\cite{lupo2020quantum} discusses the resolution limit for measuring the positions of weak point sources using an interferometric imaging system. Our paper extends this effort: we consider the resolution limit for measuring the positions of point sources of arbitrary strength using an interferometric imaging system.

As a widely-used conventional imaging method beyond single lens imaging, interferometric imaging enables an array of lenses to provide enhanced resolution compared to a single lens. Interferometric imaging is based on the Van Cittert-Zernike theorem, \cite{zernike1938concept}, which roughly speaking uses interference between signals arriving at different positions in the image plane to reconstruct the intensity distribution in the source plane. This method has led to the thriving development of interferometric telescope arrays, especially in the radio wavelength \cite{wilson2009tools,kellermann2001development}. The high angular resolution provided by the Event Horizon Telescope, a radio interferometer array, made it possible to obtain the first image of a supermassive black hole at the center of the Messier 87 Galaxy \cite{collaborat2019first1}. Recently, methods to improve interferometric imaging systems using quantum information techniques have been proposed, which show transmission loss between two nodes in an optical interferometric array can be circumvented by quantum networks \cite{gottesman2012longer,khabiboulline2019optical}. The precision with which the mutual coherence can be measured has also been explored both theoretically and experimentally using quantum estimation theory \cite{pearce2017optimal,howard2019optimal}. It is well known that the ability to resolve two point sources using interferometric telescope arrays based on the Van Cittert-Zernike theorem is limited by the longest baseline (we provide a detailed introduction to the resolution limit of interferometer arrays in Appendix \ref{conventional}). 

This resolution limit also holds even for the methods that improve the sensitivity of estimating the coherence function in Refs.~\cite{gottesman2012longer,khabiboulline2019optical,pearce2017optimal,howard2019optimal}. The reason behind this resolution limit is the incomplete sampling of the image plane, which is similar to Rayleigh's limit with finite aperture size.
Given these facts, it is then very tempting to ask whether one can achieve superresolution for interferometric imaging systems. The recent work of Ref.~\cite{lupo2020quantum} gives an affirmative answer to this question. They consider an arbitrary number of weak incoherent thermal point sources observed by a system of collectors and determine the fundamental limit of sensing the parameters related to the position of the sources. Their result shows that in the ideal case, there is no resolution limit for estimating the separation between two weak point sources using interferometric telescope arrays. 

Superresolution in interferometric imaging for the case of arbitrary intensity remains to be studied. Arbitrary intensity is important to consider because multiphoton coincidences and photon bunching, which have been ignored \cite{lupo2020quantum}, have significant effects in some situations \cite{mandel1995optical}. For the single lens case, these effects were pointed out and superresolution was shown to be achieved for two incoherent sources of arbitrary strength \cite{lupo2016ultimate,nair2016far}, and a general extended source has been considered \cite{zhou2019modern}. It should similarly be rigorously confirmed that superresolution is achievable for strong thermal sources using interferometric imaging. In addition, outside the topic of superresolution, it has been pointed out for interferometric imaging that the accuracy of estimating the coherence function using heterodyne detection exhibits very different behavior for strong versus weak thermal sources \cite{tsang2011quantum}. This is because the vacuum state dominates, which is a problem if the measurement cannot distinguish a vacuum state and a single photon state. This discussion on the estimation of the coherence function motivates us to ask the question whether superresolution of strong thermal sources also exhibits different behavior compared to weak thermal sources. In this paper, we show  using quantum estimation theory that superresolution can be achieved in interferometric imaging for thermal sources of arbitrary strength. 
Our results include the weak thermal source limit as a special case. The proposed measurement to achieve superresolution uses a linear beam splitter, which is the same as in the weak thermal source case, but here the measurement requires photon-number detection to resolve the vacuum and single photon state. We also determine the effect of misalignment on the performance.

\section*{Theoretical Model} 
\begin{figure}[!htb]
\begin{center}
\includegraphics[width=0.8\columnwidth]{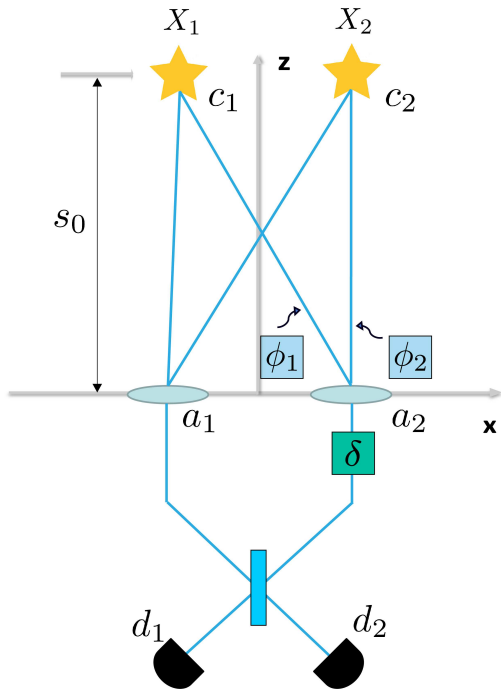}
\caption{Schematic of the setup for estimating the position of two strong thermal point sources $c_1$ and $c_2$ at positions $X_1$ and $X_2$, respectively. The light from the two sources is collected with a two-mode interferometer. States received by the two interferometer modes differ by the phases $\phi_1$ and $\phi_2$ due to the difference in path length. } 
\label{general_setting}
\end{center}
\end{figure}

We model the quantum state received from two strong thermal point sources of equal intensity by a linear interferometer with two telescopes in the paraxial regime. We assume the two point sources are incoherent, which is reasonable for astronomical observation because radiating particles from distant astronomical objects should not have any correlation \cite{goodman1985statistical}.  We assume the positions of the two point sources can be described in one dimension as $X_1$ and $X_2$, as shown in Fig.~\ref{general_setting}. The two point sources are assumed to be monochromatic and can be described by the canonical annihilation and creation operators $c_1$, $c_1^\dagger$ and $c_2$, $c_2^\dagger$. The two modes $a_1$, $a_1^\dagger$ and $a_2$, $a_2^\dagger$ of the interferometer in the image plane receive the state from the sources with phases $\phi_1$ and $\phi_2$ due to the difference in light path length, which contains information on the position of the sources. We explicitly derive the relation between $\phi_1$, $\phi_2$ and the parameters of the settings (detailed in Appendix B) as
\begin{equation}
\phi_i=kB\frac{X_i}{s_0},\quad i=1,2,
\end{equation}
where $B$ is the length of the baseline, $k$ is the wavevector of the light, and $s_0$ is the longitudinal distance to the source plane.

 Similar to the derivation in Ref. \cite{lupo2016ultimate}, the states received in the interferometer modes $a_1$ and $a_2$ are an attenuated version of the source modes $c_1$ and $c_2$:
\begin{equation}
c_i\rightarrow \sqrt{\eta}a_1+\sqrt{\eta}e^{i\phi_i}a_2+\sqrt{1-2\eta}\,v_i,\quad i=1,2,
\end{equation}
where $v_i$ are auxiliary environmental modes and $\eta$ is the attenuation ratio. Starting from the thermal states of the source $c_1$ and $c_2$, we derive the states received by the interferometer (detailed in Appendix C) as
\begin{align}\label{eq:state}
&\rho=\frac{1}{(\pi\eta\bar{N})^2}\int_{C^2}d^2\alpha_1d^2\alpha_2\exp \left(-\frac{\abs{\alpha_1}^2+\abs{\alpha_2}^2}{\eta\bar{N}}\right)\nonumber\\
&\times\left[\ket{\alpha_1+\alpha_2}\bra{\alpha_1+\alpha_2}_{a_1}\right.\\
&\otimes\ket{\alpha_1e^{-i\phi_1}+\alpha_2e^{-i\phi_2}}\bra{\alpha_1e^{-i\phi_1}+\alpha_2e^{-i\phi_2}}_{a_2}\left.\right], \nonumber
\end{align}
where $\bar{N}$ represents the strength of each source and $\ket{\alpha_1+\alpha_2}$ and ${\ket{\alpha_1e^{-i\phi_1}+\alpha_2e^{-i\phi_2}}}$ are the coherent states of the two interferometer modes $a_1$ and $a_2$.
We confirm the derived state is still a Gaussian state in Appendix C. Gaussian states are completely characterized by their mean displacement ${\lambda_\mu=\operatorname{Tr}\left[\rho \mathbf{a}_{\mu}\right]}$, where $\mathbf{a}=[a_1,a_1^\dagger,a_2,a_2^\dagger]$,  and covariance matrix ${\Sigma_{\mu \nu} =\frac{1}{2} \operatorname{Tr}\left[\rho\left(\tilde{\mathbf{a}}_{\mu} \tilde{\mathbf{a}}_{\nu}+\tilde{\mathbf{a}}_{\nu} \tilde{\mathbf{a}}_{\mu}\right)\right]}$, with ${\tilde{\mathbf{a}}_{\mu}=\mathbf{a}_{\mu}-\lambda_{\mu}}$  \cite{braunstein2005quantum,weedbrook2012gaussian}. The mean displacement $\lambda_\mu$ and covariance matrix $\Sigma$ of $\rho$ are given by
\begin{equation}
\begin{aligned}
&\lambda_\mu=0,\quad \text{for}\,\,\forall \mu, \\
&\Sigma=\left[\begin{matrix}
0 & p & 0 & q\\
p & 0 & q^* & 0\\
0 & q^* & 0 & p\\
q & 0 & p & 0
\label{eq:Sigma}
\end{matrix}\right],
\end{aligned}
\end{equation}
where ${p=2\eta\bar{N}+\frac{1}{2}}$ and  ${q=(e^{i\phi_1}+e^{i\phi_2})\eta \bar{N}}$.

\section*{Fundamental Sensitivity Limit}
We now consider the fundamental limit of resolving two point sources with two telescopes. For two point sources, the resolution is reflected in the sensitivity of measuring the centroid ${\theta_1=\frac{1}{2}(X_1+X_2)}$ and the separation between the two sources ${\theta_2=X_1-X_2}$. The sensitivity of estimating $\theta_1$ and $\theta_2$ is bounded by the quantum Fisher information (QFI) $F$: ${\Sigma_{\vec{\theta}}\geq F^{-1}}$, with its $(\mu,\nu)$ element ${[\Sigma_{\vec{\theta}}]_{\mu \nu}=\mathbb{E}\left[(\theta_\mu-\check{\theta}_\mu)(\theta_\nu-\check{\theta}_\nu)\right]}$, where $\check{\theta}_\mu$ is the unbiased estimator of the $\mu$-th unknown parameter. This sensitivity limit given by the QFI is the quantum Cram\'{e}r-Rao bound (QCRB) \cite{helstrom1976quantum}. The matrix element $F_{ij}$ of the QFI of a Gaussian state has been derived as a closed-form expression in terms of $\lambda_\mu$ and $\Sigma$ in Refs.~\cite{monras2013phase,gao2014bounds}:
\begin{equation}\label{QFI_formula}
F_{i j}=\frac{1}{2} \mathfrak{M}_{\alpha \beta, \mu \nu}^{-1} \partial_{j} \Sigma_{\alpha \beta} \partial_{i} \Sigma_{\mu \nu}+\Sigma_{\mu \nu}^{-1} \partial_{j} \lambda_{\mu} \partial_{i} \lambda_{\nu},
\end{equation}
where ${\mathfrak{M}=\Sigma \otimes \Sigma+\frac{1}{4} \Omega \otimes \Omega}$, with ${\Omega=\bigoplus_{k=1}^{n} i \sigma_{y}}$ where $\sigma_{y}$ is the Pauli $y$ matrix, $\partial_j$ is the derivative over the $j$-th unknown parameter, and repeated indices imply summation.

\begin{figure}[!htb]
\begin{center}
\includegraphics[width=1\columnwidth]{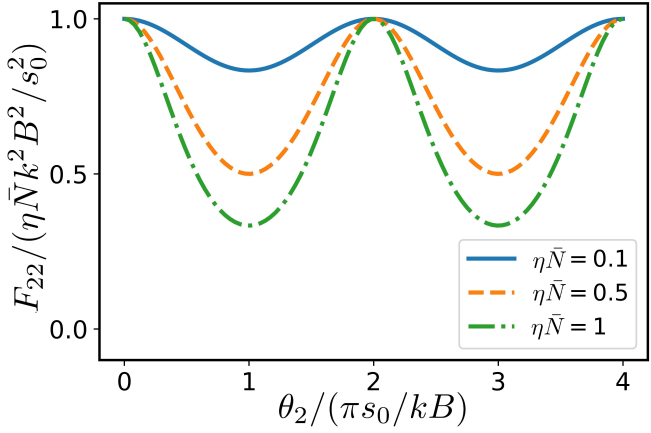}
\caption{The quantum Fisher information $F_{22}$ for estimating the separation, in units of $\eta \bar{N} k^2 B^2/s_0^2$, as a function of the separation $\theta_2$, for different source strengths $\bar{N}$.} 
\label{QFI_N_}
\end{center}
\end{figure}

The quantum Fisher information for the separation $\theta_2$ is then given by
\begin{align}\label{F22}
F_{22}&=-\frac{k^2B^2}{s_0^2}\frac{\eta \bar{N}(1+3\eta\bar{N}+\eta\bar{N}\cos(\phi_1-\phi_2))}{-1-2\eta \bar{N}(2+\eta N)+2\eta^2\bar{N}^2\cos(\phi_1-\phi_2)}\nonumber\\
&\xlongequal[]{\theta_2\rightarrow0}\frac{k^2B^2}{s_0^2}\eta\bar{N}.
\end{align}
We emphasize that when the separation between the two point sources tends to zero, i.e.,~${\theta_2\rightarrow 0}$, the quantum Fisher information tends to a constant. This implies that there is actually no resolution limit for resolving two strong thermal point sources.  Notice the QFI here is proportional to $B^2$, where $B$ is the baseline of an interferometer. Compared with a single lens, where the QFI is proportional to $D^2$ \cite{tsang2016quantum}, where $D$ is the diameter of a single lens, an interferometer has much larger QFI since $B\gg D$. 
  The quantum Fisher information $F_{11}$ for estimating the centroid, $\theta_1$, also tends to a constant as the separation ${\theta_2\rightarrow 0}$, as detailed in Appendix D; we discuss this result after analyzing the QFI for the separation. We plot the values of $F_{22}$ as a function of the separation $\theta_2$ for different source strengths $\bar{N}$ in Fig.~\ref{QFI_N_}. The QFI shows periodicity over $\theta_2$ with period $2\pi s_0/(kB)$, which is roughly the conventional resolution limit of interferometry. The periodicity is due to the fact that the position information of the sources is encoded in the phase $e^{i\phi_j}$. Adding $2\pi$ to $\phi_j$ does not affect the state described in Eq.~\ref{eq:state} and hence cannot be distinguished by any measurement. The setup considered in our model can only distinguish  $\phi_1-\phi_2\propto X_1-X_2$ up to an integer number of $2\pi$. We can solve this problem by having detectors at more than two positions.  This is different from direct imaging, where varying the positions of the sources will always affect the received states and hence there is no periodicity in the QFI  \cite{lupo2016ultimate,nair2016far}. We observe that for the intermediate values of $\theta_2$ within a period, the QFI decreases with increasing source strength $\bar{N}$; this is in contrast to the limit of a weak thermal source, in which the QFI is a constant versus separation $\theta_2$ \cite{lupo2020quantum}. As pointed out by Ref.~\cite{nair2016far} in the single lens case, this is a net result of multiphoton events. Note that Fig.~\ref{QFI_N_} is a plot of the QFI \textit{per photon}, which decreases for some values of $\theta_2$ as source strength increases, but a stronger source still has a larger total QFI of estimating $\theta_2$ given its larger photon number. We have  verified this and  found $\partial F_{22}/\partial( \eta \bar{N})$ is always positive for all possible parameters. 

We now consider what the proper measurement strategy is to actually achieve this limit.  The positive operator-valued measure (POVM) that can saturate the QCRB is given by the eigenbasis of the symmetric logarithmic derivative (SLD) \cite{braunstein1994statistical,paris2009quantum}. For a Gaussian state, the SLD has been derived in terms of its mean displacement $\lambda_\mu$ and covariance matrix $\Sigma$ \cite{monras2013phase,gao2014bounds}:  
\begin{equation}\label{SLD_formula}
\mathcal{L}_{i}=\frac{1}{2} \mathfrak{M}_{\gamma \kappa, \alpha \beta}^{-1}\left(\partial_{i} \Sigma_{\alpha \beta}\right)\left(a_{\gamma} a_{\kappa}-\Sigma_{\gamma \kappa}\right),
\end{equation}
where $a_i$ is the mode operator  and we sum over repeated indices.
The SLD for estimating the separation $\theta_2$ is:
\begin{equation}
\mathcal{L}_{\theta_2}=2l_1a^\dagger_1a_1+2l_1a_2^\dagger a_2+2l_2a_1a_2^\dagger+2l_2^*a_1^\dagger a_2+C_{\theta_2},
\end{equation}
where
\begin{align}
&C_{\theta_2}=-\eta\bar{N}[8l_1+2l_2(e^{i\phi_1}+e^{i\phi_2})+2l_2^*(e^{-i\phi_1}+e^{-i\phi_2})],\nonumber\\
&l_1=\frac{kB}{s_0}\frac{(1+4\eta\bar{N})\cot \frac{\phi_1-\phi_2}{2}}{-4[1+2\eta\bar{N}(2+\eta\bar{N})]+8\eta^2\bar{N}^2\cos(\phi_1-\phi_2)},\\
&l_2=-\frac{kB}{s_0}\nonumber\\
&\;\;\times\frac{e^{-\frac{1}{2}i(\phi_1+\phi_2)}(1+3\eta\bar{N}+\eta\bar{N}\cos(\phi_1-\phi_2))\csc\frac{\phi_1-\phi_2}{2}}{4[-1-2\eta\bar{N}(2+\eta\bar{N})+2\eta^2\bar{N}^2\cos(\phi_1-\phi_2)]}.\nonumber
\end{align}

To find the eigenbasis of the SLD, we diagonalize  $\mathcal{L}_{\theta_2}$. Assuming ${d_1=\frac{1}{\sqrt{2}}(a_1+e^{i\delta}a_2)}$, ${d_2=\frac{1}{\sqrt{2}}(a_1-e^{i\delta}a_2)}$ and dropping the constant terms, we have
\begin{align}
\mathcal{L}_{\theta_2}=&(2l_1+l_2e^{i\delta}+l_2^*e^{-i\delta})d_1^\dagger d_1\nonumber\\
&+(2l_1-l_2e^{i\delta}-l_2^*e^{-i\delta})d_2^\dagger d_2\\
&+(l_2e^{i\delta}-l_2^*e^{-i\delta})d_1^\dagger d_2-(l_2e^{i\delta}-l_2^*e^{-i\delta})d_2^\dagger d_1.\nonumber
\end{align}
We can choose ${l_2e^{i\delta}-l_2^*e^{-i\delta}=0}$ or equivalently ${\delta=\frac{1}{2}(\phi_1+\phi_2)}$, which means the SLD has the Fock basis of $d_1$, $d_2$ as its eigenbasis.  Thus, the optimal POVM for estimating $\theta_2$ is $\{\ket{m,n}_d\bra{m,n}_{d}\}_{\{m,n\}}$, with ${d_1^\dagger d_1\ket{m,n}_d=m\ket{m,n}_d}$ and ${d_2^\dagger d_2\ket{m,n}_d=n\ket{m,n}_d}$. 

\begin{figure}[!htb]
\begin{center}
\includegraphics[width=1\columnwidth]{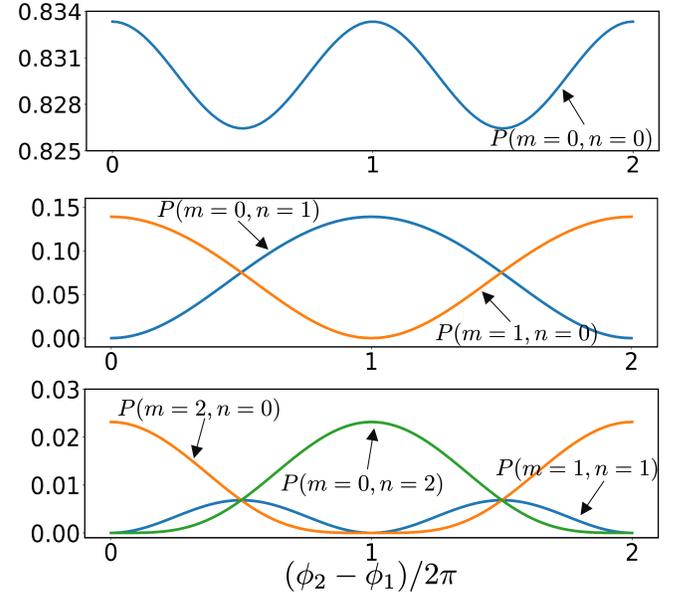}
\caption{The probability $P(m,n)$ of projecting the state onto $\ket{m,n}_d$ with $\eta \bar{N}=0.1$ as a function of $\phi_2-\phi_1$.} 
\label{Pmn}
\end{center}
\end{figure}

\begin{figure}[!htb]
\begin{center}
\includegraphics[width=1\columnwidth]{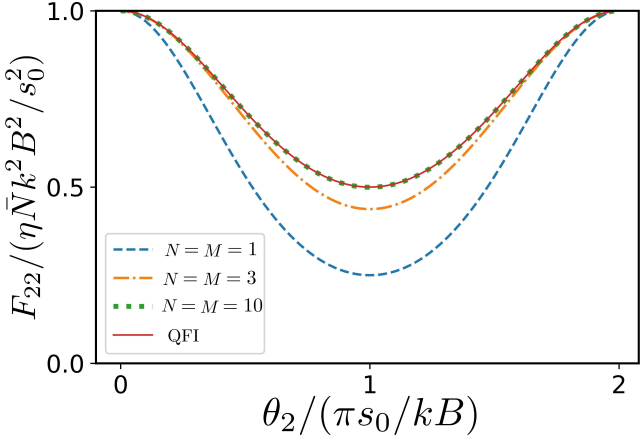}
\caption{Fisher information for photon number detection that only distinguishes the Fock state $\ket{m,n}$ for $m\leq M$, $n\leq N$. Events with greater photon number are ignored. } 
\label{FI}
\end{center}
\end{figure}

As shown in Fig.~\ref{general_setting}, we can implement the above POVM by combining the states of the two modes of the two telescopes on a beam splitter, adding a fixed phase delay $\delta$ corresponding to the optimal delay found above to one of the arms, and performing photon-number-resolved detection in both of the two output ports. This setup is the same as found in Ref.~\cite{lupo2020quantum}, except for the photon-number-resolved detection. More specifically, for the weak thermal source discussed in  Ref.~\cite{lupo2020quantum} and in Appendix E, the quantum state $\rho$ 
received by the two telescopes in modes $a_1$ and $a_2$ is in a Hilbert space spanned by the Fock state basis $\{\ket{m,n}\}$ with constraint $m+n\le 1$ -- no such constraint is present for the case of a strong thermal source. In order to implement the POVM found above, the state is measured for each temporal mode and projected onto one of the Fock bases $\ket{m,n}_d$ of $d_{1,2}$ modes.  Data are accumulated to find the probability $P(m,n)$ of getting each outcome. The probability distribution is then fit with its corresponding theoretical prediction to obtain the unknown centroid or separation. The theoretical prediction for the first few $P(m,n)$ is given in Fig.~\ref{Pmn}. The probability distribution $P(m,n)$ is symmetric with respect to $\phi_2-\phi_1=2\pi$ as a function of $\phi_2-\phi_1$, so for some separations there exists ambiguity in the estimation -- this is resolved through measurement of the centroid $\theta_1\propto\phi_1+\phi_2$, discussed below. In practice, the detectors may be able to distinguish only the first few Fock states of low photon number. As shown in Fig.~\ref{FI},  even if only Fock states $\ket{m,n}_d$ with ${m\leq M}$, ${n\leq N}$ can be distinguished,  the FI still maintains a reasonable amount of the QFI, which implies that we can achieve a large part of the sensitivity predicted in the ideal case. In particular, we emphasize that even if we only distinguish the presence or absence of the photon, i.e. $M=N=1$, superresolution is still achieved, as is indicated by the finite FI in this case when the separation $\theta_2$ goes to zero.

We estimate the improvement our method provides compared to the conventional imaging method based on the Van Cittert-Zernike theorem using parameters similar to real present-day interferometer arrays in Appendix \ref{compareconventional}. We consider the case where the observation is made with wavelength ${\lambda=5}$~mm and longest baseline ${B=10}$~km. The resolution of the conventional method is then ${\lambda/B=5\times 10^{-7} \mathrm{radians}\approx 0.1''}$. When the angular separation of the two point sources is ${\theta_2/s_0=0.01''}$ and ${\eta \bar{N}=0.01}$, the Fisher information of our optimal measurement is larger than the conventional method by a factor of roughly 30. If we assume the mean square error of estimating the angular separation $\theta_2/s_0$ scales with the number of samples $n$ as ${\Delta (\theta_2/s_0)^2\propto 1/n}$,  this implies that our optimal measurement can shorten the observation time by a factor of 30 to achieve the same sensitivity.

\begin{figure*}[!htb]
\begin{center}
\includegraphics[width=2\columnwidth]{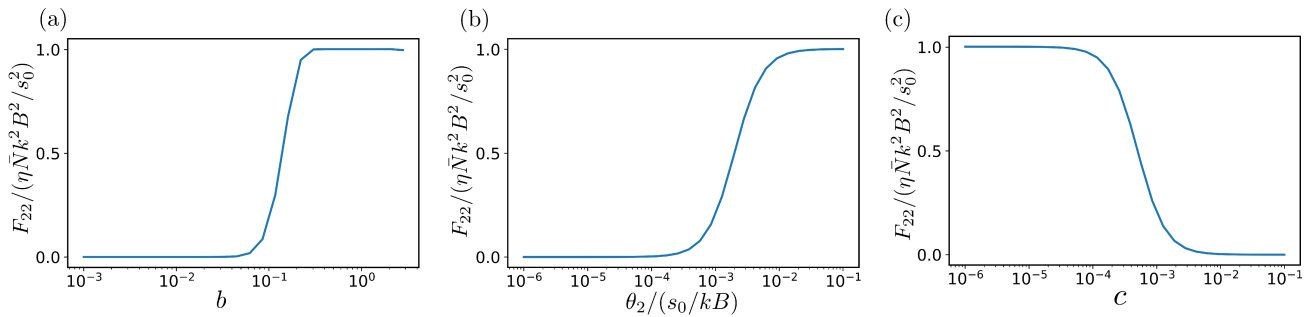}
\caption{(a) Approximate Fisher information as a function of the integration range $[0,b]$. Other parameters are chosen as $\bar{N}=0.01$; $m,n\leq 3$; $c=10^{-6}$; and $\theta_2=10^{-3}$.  (b) Approximate Fisher information  as a function of separation $\theta_2$ for fixed misalignment $c=10^{-3}$. Other parameters are chosen as $\bar{N}=0.01$ and $m,n\leq 3$. (c) Approximate Fisher information as a function of misalignment $c$ for fixed separation $\theta_2=10^{-3}$. Other parameters are chosen as $\bar{N}=0.01$ and $m,n\leq 3$.} 
\label{misalignment_combine}
\end{center}
\end{figure*}

The phase delay $\delta$ depends on the centroid $\theta_1$ of the two point sources, since ${\theta_1\propto\phi_1+\phi_2}$, and thus the scheme requires accurate measurement of the centroid. The derivation for the estimation of the centroid is detailed in Appendix D; there it can been seen that a similar measurement strategy that also depends on the centroid is optimal. This recursive relationship can be overcome, as in conventional imaging there is no fundamental limitation on the accuracy of estimating the centroid; thus, other imaging methods can be used or suboptimal strategies can be constructed to determine the centroid, such as using a random phase scheme.  Nevertheless, misalignment of the centroid must be taken into account. We now show how the superresolution predicted by the QFI is affected by a deviation of $\delta$ from $\frac{1}{2}(\phi_1+\phi_2)$. We write the deviation as $c=\frac{1}{2}(\phi_1+\phi_2)-\delta$. We write the state $\ket{m,n}_d$ using the Fock basis of modes $a_1$, $a_2$:
\begin{equation}
\begin{aligned}
\ket{m,n}_d&=(d_1^\dagger)^m(d_2^\dagger)^n\ket{0}\\
&=2^{-\frac{m+n}{2}}(a_1^\dagger+e^{-i\delta}a_2^\dagger)^m(a_1^\dagger-e^{-i\delta}a_2^\dagger)^n\ket{0}\\
&=2^{-\frac{m+n}{2}}\sum_{j,k}C_m^jC_n^k(-1)^ke^{-i(j+k)\delta}\\
&\quad\quad\quad\quad\quad\quad\times(a_1^\dagger)^{m+n-j-k}(a_2^\dagger)^{j+k}\ket{0},
\end{aligned}
\end{equation}
where $C_m^j=\frac{m!}{j!(m-j)!}$. We then evaluate
\begin{equation}
\begin{aligned}
&f(m,n,\alpha_1,\alpha_2)\\
&= \bra{m,n}_d\left(\ket{\alpha_1+\alpha_2}\otimes\ket{\alpha_1 e^{-i\phi_1}+\alpha_2 e^{-i\phi_2}}\right)\\
&=2^{-\frac{m+n}{2}}\sum_{j,k}C_m^j C_n^k(-1)^ke^{i(j+k)\delta}\\
&\quad\quad\quad\times e^{-\frac{1}{2}|\alpha_1+\alpha_2|^2-\frac{1}{2}|\alpha_1e^{-i\phi_1}+\alpha_2e^{-i\phi_2}|^2}\\
&\quad\quad\quad\times(\alpha_1+\alpha_2)^{m+n-j-k}(\alpha_1 e^{-i\phi_1}+\alpha_2 e^{-i\phi_2})^{j+k}.
\end{aligned}
\end{equation}
The probability of getting outcome $\ket{m,n}_d\bra{m,n}_d$ is given by 
\begin{equation}\begin{aligned}
P_d(m,n)&=\frac{1}{(\pi\bar{N})^2}\int_{C^2}d^2\alpha_1d^2\alpha_2\\
&\times\exp \left(-\frac{\abs{\alpha_1}^2+\abs{\alpha_2}^2}{\bar{N}}\right) |f(m,n,\alpha_1,\alpha_2)|^2.
\end{aligned}\end{equation}
The Fisher information of estimating the separation is calculated as 
\begin{equation}\begin{aligned}
FI=\sum_{m,n=0}^\infty \frac{(\partial P_d(m,n)/\partial \theta_2)^2}{P_d(m,n)}.
\end{aligned}\end{equation}
This calculation is intractable both analytically and numerically. We  instead make the following approximation. First, we only keep the contribution of $\ket{m,n}_d\bra{m,n}_d$ with $m\leq 3$, $n\leq 3$. As pointed out above, keeping only the first few elements of the POVM can still achieve superresolution; i.e., the Fisher information tends to a constant when the separation $\theta_2\rightarrow 0$. Secondly, we do the integration for the phase and amplitude of $\alpha_1$, $\alpha_2$ separately and define a cut-off for the integration of the amplitude:
\begin{equation}\begin{aligned}
P_d(m,n)&\approx \frac{1}{(\pi\bar{N})^2} \int_0^bd|\alpha_1|\int_0^bd|\alpha_2||\alpha_1||\alpha_2|\\
&\times\exp \left(-\frac{\abs{\alpha_1}^2+\abs{\alpha_2}^2}{\bar{N}}\right)g(m,n,|\alpha_1|,|\alpha_2|),
\end{aligned}\end{equation}
\begin{equation}\begin{aligned}
g(m,n,|\alpha_1|,|\alpha_2|)&=\int_0^{2\pi}d\beta_1\int_0^{2\pi}d\beta_2\, \\
&\times |f(m,n,|\alpha_1|e^{i\beta_1},|\alpha_2|e^{i\beta_2})|^2,
\end{aligned}\end{equation}
where $\alpha_1=|\alpha_1|e^{i\beta_1}$, $\alpha_2=|\alpha_2|e^{i\beta_2}$, and $b$ is a finite number to introduce a cutoff for the integral for convenience in the numerical calculation. For a fixed value of $\bar{N}$, the FI tends to a constant value as $b$ increases, as shown in Fig.~\ref{misalignment_combine}(a), which validates the cutoff. 

 We plot the FI as a function of separation $\theta_2$ with fixed misalignment $c$ in Fig.~\ref{misalignment_combine}(b). It is clear that, with a nonzero misalignment, when the separation tends to zero the FI vanishes and superresolution cannot be achieved. We also plot the FI as a function of the misalignment $c$ with fixed separation $\theta_2$ in Fig.~\ref{misalignment_combine}(c). We observe that increasing the misalignment significantly degrades the FI. The threshold is roughly $c\approx \theta_2$ from the figure.

 We emphasize that even though the FI is no longer constant as $\theta_2\rightarrow 0$ in the presence of misalignment, it is still possible to get some benefit from our measurement if the misalignment $c$ is small enough compared to the separation $\theta_2$. For example, when $c=10^{-3}$ and $\theta_2/(s_0/kB)=10^{-1}$, the FI of our measurement approaches $\eta\bar{N}k^2B^2/s_0^2$ [from Fig.~\ref{misalignment_combine}(b)], while for the conventional method, the FI is $2\times 10^{-3}$ times smaller (from Fig.~\ref{misalignment_compare}). Thus, when working below the resolution limit of the conventional method ($\theta_2<s_0/kB$), our method can still significantly outperform the conventional one if the misalignment is not too large. This behavior is similar to what was found for a single lens in the presence of misalignment \cite{tsang2016quantum}: as long as the misalignment is small enough, the FI of estimating the separation is better than the direct imaging method.

\textit{Conclusion}- In summary, we have used quantum estimation theory to determine the fundamental limit of resolving two identical thermal point sources of any strength. The results show that, unlike the conventional imaging method based on the Van Cittert-Zernike theorem, a more properly designed measurement scheme can achieve a resolution not limited by the longest baseline. We find a measurement scheme using a beam splitter and photon-number-resolving detection can achieve the resolution given by the quantum Cram\'{e}r-Rao bound.   This paper can be extended to several other cases, such as resolving two point sources of unequal strength \cite{vrehavcek2017multiparameter,vrehavcek2018optimal}, estimating separation in three dimensions similarly to Refs.~\cite{yu2018quantum,lupo2020quantum}, and imaging a general extended source similarly to Refs.~\cite{zhou2019modern,tsang2019quantum}. Although we are unable to find an analytical solution for the case of multiple sources and detectors, it is at least possible to numerically calculate the QFI and SLD following a similar procedure to that briefly discussed in Appendix \ref{multiple}. As in single-lens imaging with noisy detectors \cite{lupo2020subwavelength,llen2020resolution}, we expect the signal-to-noise ratio to limit the resolution.  We hope our result inspires more discussion along these lines.

\section*{Acknowledgements}
 We thank Offir Cohen, Andrew Jordan, Eric Chitambar, Paul Kwiat, John D. Monnier, Shayan Mookherjea, Michael G. Raymer, Brian J. Smith, Robert Czupryniak, John Steinmetz and Jing Yang for helpful discussion.
This work was supported by the multi-university National Science Foundation Grant No.~1936321 -- QII-TAQS: Quantum-Enhanced Telescopy.

\onecolumngrid
\appendix
\renewcommand\thefigure{\thesection\arabic{figure}}
\setcounter{figure}{0}   

\section{Resolution limit of the conventional method}\label{conventional}
Here we briefly review the resolution limit of the conventional imaging method based on the Van Cittert-Zernike theorem. We refer the reader to Ref.~\cite{wilson2009tools,kellermann2001development} for more details. The Van Cittert-Zernike theorem relies on the fact that the mutual coherence function of the signal $V(u,v)$ between two points on the image plane is the Fourier transformation of the intensity distribution $I(l,m)$ in the source plane:
\begin{equation}
V(u,v)=\int\int I(l,m)\exp[2\pi i(lu+mv)]dldm,
\end{equation}
where $(u,v)$ are the coordinates of the baseline between the two observation points in the image plane, and $(l,m)$ are the coordinates of one point in the source plane. Of course, if we could measure all the Fourier components, we could completely reconstruct the intensity distribution, i.e. the image, with no resolution limit. But this requires us to measure the entire function $V(u,v)$, with each point of this function obtained by a measurement with particular baseline $\vec{B}=(u,v)$, which is practically impossible. 

We now determine the resolution for a finite number of samples of the image plane. We first introduce a sampling function $S(u,v)$ that takes value $S(u,v)=1$ at the points we measure and takes value $S(u,v)=0$ where we do not measure, for simplicity. We then define its Fourier transformation $B(l,m)=FT\{S(u,v)\}$, where $FT\{\cdot\}$ means Fourier transformation; this is similar to the point spread function (PSF) in the single lens imaging method, usually called the dirty beam. Performing an inverse Fourier transformation on the measured coherence function gives $I^D(l,m)=FT\{V(u,v)S(u,v)\}=I'(l,m)*B(l,m)$, where $*$ means convolution and $I'(l,m) = FT\{V(u,v)\}$ is the actual intensity distribution; $I^D(l,m)$ is usually called the dirty image. Mathematically, it is not possible to take the inverse of the convolution. In astronomical observation, a de-convolution method is carefully designed to gain some information from the dirty image, but as the convolution is not invertible, these empirical methods rely on some assumptions and provide limited resolution that depends on the length of the baseline. We could say the resolution is limited by the dirty beam $B(l,m)$ in this method, which is very similar to the resolution limit of conventional imaging systems with a single lens, where the PSF causes the limitation due to the finite size of the aperture.  So, we might intuitively regard $B(l,m)$ as an effective PSF. Looking at it another way, for the single lens case, the state before the light passes through the lens corresponds to the Fourier component of the image, so we can roughly say that a Fourier transformation is done to the state by passing through the lens, which introduces the PSF. If the lens is infinitely large, we get all the Fourier components and hence the resolution is infinitely good.

As a simple example, consider the sampling function $S(u,v)=1$ if $-d\leq u\leq d$ and $-d\leq v\leq d$ and $S(u,v)=0$ everywhere else, which means the longest baseline is $d$. Then the dirty beam is ${B(l,m)=FT\{S(u,v)\}=\frac{4}{\pi^2}\frac{\sin ld}{l}\frac{\sin md}{m}}$, and the width of it in each direction is $\pi/d$. From this we see the resolution of the interferometer array is roughly determined by its longest baseline. Of course, a real sampling function does not have this simple form; the dirty beam has structure rather than looking like a point. A deconvolution is thus usually necessary to remove the structure introduced by the dirty beam. 

We emphasize that whenever imaging is based on the Van Cittert-Zernike theorem, the resolution is limited by the effect of finite sampling. In this sense, all the discussions on improving the estimation of the coherence function in Refs.~\cite{gottesman2012longer,khabiboulline2019optical,pearce2017optimal,howard2019optimal} have this resolution limit.

\section{Relation between phase and the position of the source}

\begin{figure}[!htb]
\begin{center}
\includegraphics[width=0.5\columnwidth]{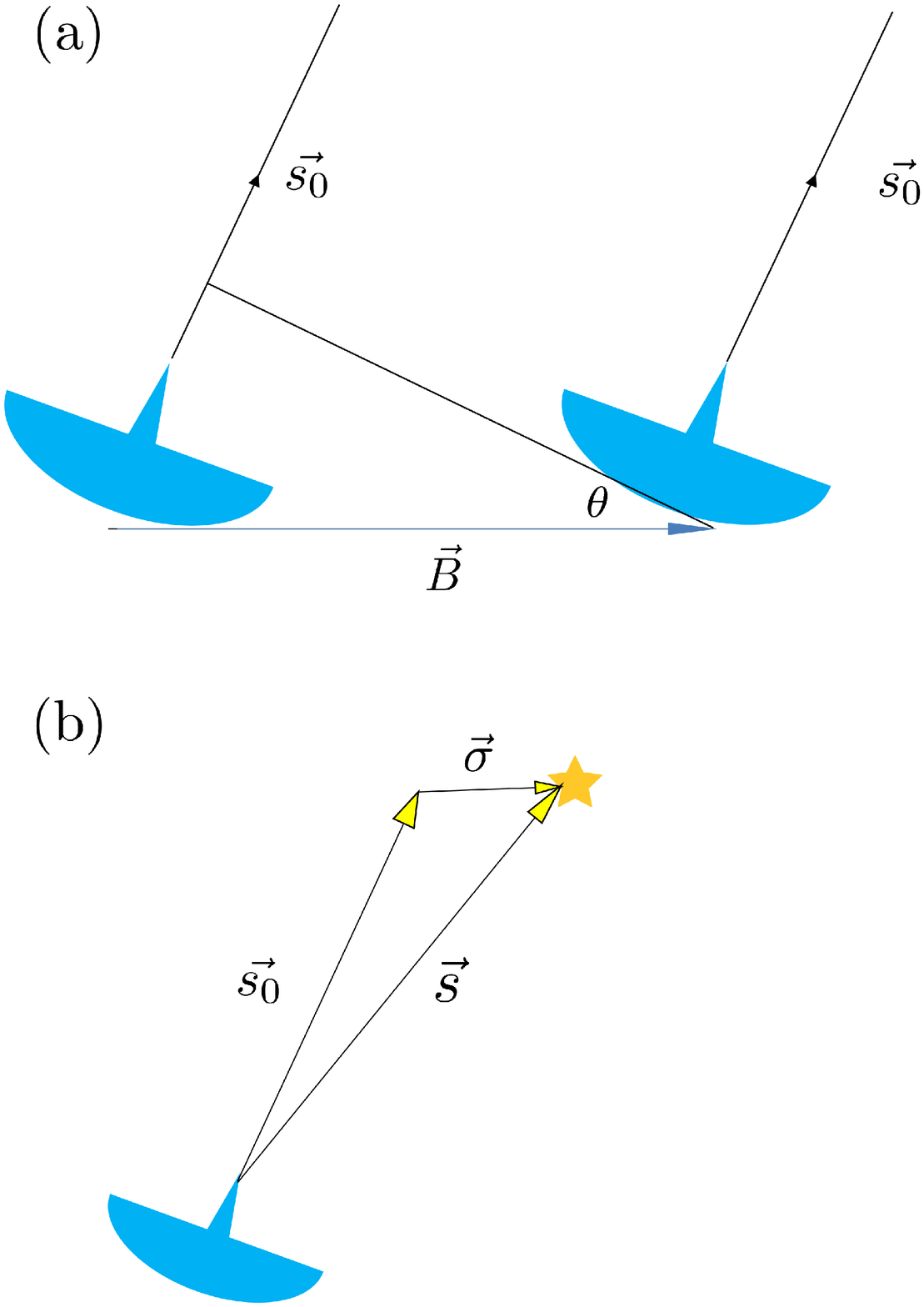}
\caption{Diagram defining the position of the sources relative to the telescopes. (a) $\vec{B}$ is a vector representing the baseline and $\theta$ is the angle of the image plane relative to the source plane. $\vec{s}_0$ is a vector connecting the observation point to the point on the source plane at which the telescope is pointing. (b) $\vec{s}$ is a point on the source and $\vec{\sigma}=\vec{s}-\vec{s}_0$. The phase (time) delay of signals received on the two telescopes encodes information on position.} 
\label{phase}
\end{center}
\end{figure}

Here we derive the relation between the phases $\phi_1$ and $\phi_2$ and the positions of the sources. As shown in Fig.~\ref{phase}, we assume the telescopes are pointing at a point $\vec{s}_0$ on the source plane. The relative positions of the two point sources are ${\vec{\sigma}_1=(X_1,0)}$ and ${\vec{\sigma}_2=(X_2,0)}$ and thus the positions of the point sources are ${\vec{s}_1=\vec{s}_0+\vec{\sigma}_1}$ and ${\vec{s}_2=\vec{s}_0+\vec{\sigma}_2}$. The phase differences $\phi_1$ and $\phi_2$ between the light arriving from point sources $X_1$ and $X_2$ at the telescopes are then
\begin{equation}
\begin{aligned}
\phi_1&=k\frac{\vec{B}\cdot\vec{s}_1}{\abs{\vec{s}_1}}=(kBs_0\sin\theta+kBX_1\cos\theta)/\abs{\vec{s}_1},\\
&=kB\sin\theta+kB\cos\theta\frac{X_1}{s_0}+o\left(\frac{X_1}{s_0}\right),
\end{aligned}
\end{equation}
\begin{equation}
\begin{aligned}
\phi_2&=k\frac{\vec{B}\cdot\vec{s}_2}{\abs{\vec{s}_2}}=(kBs_0\sin\theta+kBX_2\cos\theta)/\abs{\vec{s}_2}\\
&=kB\sin\theta+kB\cos\theta\frac{X_2}{s_0}+o\left(\frac{X_2}{s_0}\right),
\end{aligned}
\end{equation}
where we have assumed $X_1,X_2\ll s_0$ and expanded the phase as a series in $\frac{X_i}{s_0}$; the little o notation $o(\cdot)$ means the remaining terms are of order smaller than the terms in parentheses. In the main text, we assume the image plane is parallel to the source plane for simplicity; i.e., $\theta=0$, because a nonvanishing $\theta$ shows no effect on the conclusion.

\section{States received in modes \texorpdfstring{$a_1$}{a1} and \texorpdfstring{$a_2$}{a2}}
We assume the states emitted by the two point sources are thermal states $\rho^{th}$ with mean photon number $\bar{N}$. The thermal states of two modes $c_1$ and $c_2$ corresponding to the two point sources are described in Ref.  \cite{weedbrook2012gaussian} as
\begin{equation}
\begin{aligned}
\rho=\rho^{th}(\bar{N})\otimes\rho^{th}(\bar{N})
=\frac{1}{(\pi\bar{N})^2}\int_{C^2}d^2\alpha_1d^2\alpha_2\exp \left(-\frac{\abs{\alpha_1}^2+\abs{\alpha_2}^2}{\bar{N}}\right)\ket{\alpha_1}\bra{\alpha_1}_{c_1}\otimes\ket{\alpha_2}\bra{\alpha_2}_{c_2},
\end{aligned}
\end{equation}
where $\ket{\alpha_1}$ and $\ket{\alpha_2}$ are the coherent states of $c_1$, $c_2$. Using the transformation from $c_1$, $c_2$ to the modes of telescopes \cite{lupo2016ultimate},  \begin{equation}
\begin{aligned}
&c_1\rightarrow \sqrt{\eta}a_1+\sqrt{\eta}e^{i\phi_1}a_2+\sqrt{1-2\eta}\,v_1,\\
&c_2\rightarrow \sqrt{\eta}a_1+\sqrt{\eta}e^{i\phi_2}a_2+\sqrt{1-2\eta}\,v_2,
\end{aligned}
\end{equation}
We can regard the lossy process as an unbalanced multiport beam splitter. Assuming only the sources $c_{1,2}$ are radiating photons, the transformation of other input modes is irrelevant to our model. For example, if source $c_1$ generates a photon, described as $c_1^\dagger\ket{0}$, and the photon passes through an unbalanced multiport beam splitter, we obtain a coherent superposition of the three output ports $\sqrt{\eta}a_1^\dagger\ket{0}+\sqrt{\eta}e^{-i\phi_1}a_2^\dagger\ket{0}+\sqrt{1-2\eta}v_1^\dagger\ket{0}$.  We derive the state of $a_{1,2}$ analogously below, where $c_{1,2}$ are thermal states.
\begin{equation}
\ket{\alpha_1}_{c_1}\otimes\ket{\alpha_2}_{c_2}=D(\alpha_1)D(\alpha_2)\ket{0}\otimes\ket{0},
\end{equation}
\begin{equation}
\begin{aligned}
D(\alpha_1)D(\alpha_2)&=\exp(\alpha_1 c_1^\dagger-\alpha_1^* c_1)\exp(\alpha_2 c_2^\dagger-\alpha_2^* c_2)\\
=&\exp(\frac{1}{2}\eta\alpha_1\alpha_2^*-\frac{1}{2}\eta \alpha_1^*\alpha_2)\exp(\frac{1}{2}\eta\alpha_1\alpha_2^*e^{i(\phi_2-\phi_1)}-\frac{1}{2}\eta \alpha_1^*\alpha_2e^{i(\phi_1-\phi_2)})\\
&\times\exp [\sqrt{\eta}(\alpha_1+\alpha_2)a_1^\dagger-\sqrt{\eta}(\alpha_1^*+\alpha_2^*)a_1]\exp [\sqrt{\eta}(\alpha_1e^{-i\phi_1}+\alpha_2e^{-i\phi_2})a_2^\dagger-\sqrt{\eta}(\alpha_1^*e^{i\phi_1}+\alpha_2^*e^{i\phi_2})a_2]\\
&\times\exp[\sqrt{1-2\eta}\alpha_1v_1^\dagger-\sqrt{1-2\eta}\alpha_1^*v_1]\exp[\sqrt{1-2\eta}\alpha_2v_2^\dagger-\sqrt{1-2\eta}\alpha_2^*v_2].
\end{aligned}
\end{equation}
Hence, the state evolves to be
\begin{equation}
\begin{aligned}
&\ket{\alpha_1}_{c_1}\otimes\ket{\alpha_2}_{c_2}\rightarrow C\ket{\sqrt{\eta}(\alpha_1+\alpha_2)}_{a_1}\otimes\ket{\sqrt{\eta}(\alpha_1e^{-\phi_1}+\alpha_2e^{-i\phi_2})}_{a_2}\otimes\ket{\sqrt{1-2\eta}\alpha_1}_{v_1}\otimes\ket{\sqrt{1-2\eta}\alpha_2}_{v_2},\\
&C=\exp(\frac{1}{2}\eta\alpha_1\alpha_2^*-\frac{1}{2}\eta \alpha_1^*\alpha_2)\exp(\frac{1}{2}\eta\alpha_1\alpha_2^*e^{i(\phi_2-\phi_1)}-\frac{1}{2}\eta \alpha_1^*\alpha_2e^{i(\phi_1-\phi_2)}).
\end{aligned}
\end{equation}
We then find the state received by the two modes of the two telescopes $a_1$, $a_2$ as
\begin{equation}
\begin{aligned}
\rho\rightarrow\frac{1}{(\pi\eta\bar{N})^2}\int_{C^2}d^2\alpha_1d^2\alpha_2&\exp \left(-\frac{\abs{\alpha_1}^2+\abs{\alpha_2}^2}{\eta\bar{N}}\right)\ket{\alpha_1+\alpha_2}\bra{\alpha_1+\alpha_2}_{a_1}\otimes\ket{\alpha_1e^{-i\phi_1}+\alpha_2e^{-i\phi_2}}\bra{\alpha_1e^{-i\phi_1}+\alpha_2e^{-i\phi_2}}_{a_2}\\
&\otimes\ket{\sqrt{1/\eta-2}\alpha_1}\bra{\sqrt{1/\eta-2}-\alpha_1}_{v_1}\otimes\ket{\sqrt{1/\eta-2}\alpha_2}\bra{\sqrt{1/\eta-2}-\alpha_2}_{v_2}.
\end{aligned}
\end{equation}
We can trace out the states of the environmental modes $v_1$ and $v_2$, which we do not have access to in the measurement. We will then get
\begin{equation}
\begin{aligned}
\rho\rightarrow\frac{1}{(\pi\eta\bar{N})^2}\int_{C^2}d^2\alpha_1d^2\alpha_2&\exp \left(-\frac{\abs{\alpha_1}^2+\abs{\alpha_2}^2}{\eta\bar{N}}\right)\ket{\alpha_1+\alpha_2}\bra{\alpha_1+\alpha_2}_{a_1}\otimes\ket{\alpha_1e^{-i\phi_1}+\alpha_2e^{-i\phi_2}}\bra{\alpha_1e^{-i\phi_1}+\alpha_2e^{-i\phi_2}}_{a_2}.
\end{aligned}
\end{equation}

We now verify the derived state of $a_1$, $a_2$ above is a Gaussian state. By definition, Gaussian states written in the Wigner representation should be Gaussian, which requires the Wigner-Weyl operator to be:
\begin{equation}
\chi(\xi):=\tr{\rho e^{-\vec{a}^T\Omega\vec{\xi}}}=\exp[\frac{1}{2}\vec{\xi}^T(\Omega^T\Sigma\Omega)\vec{\xi}+\vec{\xi}\Omega\vec{\lambda}],
\label{eq:wely}
\end{equation}
where $\vec{\xi}=\{\xi_1,\xi_1^{*},\xi_2,\xi_2^{*}\}$ and $\Omega=-\sigma_y\otimes\sigma_y$ with $\sigma_y$ being the Pauli $y$ matrix. 
\par
Inserting a state of the form of Eq.~\ref{eq:state} into the left-hand side of Eq.~\ref{eq:wely}, we have:
\begin{align}
\Tr{\rho e^{-\vec{a}^T\Omega\vec{\xi}}}=&\frac{1}{(\pi\bar{N})^2}\int_{C^2}d^2\alpha_1d^2\alpha_2\exp \left(-\frac{\abs{\alpha_1}^2+\abs{\alpha_2}^2}{\bar{N}}\right)\exp\left(-\frac{|\xi_1|^2}{2}\right)\exp\left(-\sqrt{\eta}[(\alpha_1^*+\alpha_2^*)\xi_1-(\alpha_1+\alpha_2)\xi_1^*]\right)\notag\\
\times&\exp\left(-\frac{|\xi_2|^2}{2}\right)\exp\left(-\sqrt{\eta}[(\alpha_1^*e^{i\phi_1}+\alpha_2^*r^{i\phi_2})\xi_1-(\alpha_1e^{-i\phi_1}+\alpha_2^{-i\phi_2})\xi_2^*]\right)\notag\\
=&\exp[\frac{1}{2}\vec{\xi}^T(\Omega^T\Sigma\Omega)\vec{\xi}+\vec{\xi}\Omega\vec{\lambda}],
\end{align}
with $\Sigma$ given exactly as in Eq.~\ref{eq:Sigma}.

\section{Estimation of the centroid}
The QFI for the estimation of the centroid is given by
\begin{equation}
\begin{aligned}
F_{11}&=-\frac{2k^2B^2}{s_0^2}\frac{\eta \bar{N}(1+\cos(\phi_1-\phi_2))}{-1-\eta \bar{N}+\eta\bar{N}\cos(\phi_1-\phi_2)}\\
&\xlongequal[]{\theta_2\rightarrow0}4\frac{k^2B^2}{s_0^2}\eta\bar{N}.
\end{aligned}
\end{equation}
We have checked that the off-diagonal elements of the QFI vanish; i.e., ${F_{12}=F_{21}=0}$. The compatibility of optimally measuring several parameters is highly nontrivial \cite{ragy2016compatibility,chrostowski2017super}. It is hard to saturate the QCRB of estimating the centroid and the separation at the same time. We will see in the following that the optimal measurements for estimating the centroid and separation are different. To find the optimal POVM that can achieve the accuracy predicted by the QCRB, we calculate the SLD for estimating the centroid $\theta_1$ as
\begin{equation}
    \mathcal{L}_{\theta_1}=2l_3a_1a_2^\dagger+2l_3^*a_1^\dagger a_2+C_{\theta_1},
\end{equation}
where
\begin{equation}
\begin{aligned}
C_{\theta_1}&=-\eta\bar{N}[2l_3(e^{i\phi_1}+e^{i\phi_2})+2l_3^*(e^{-i\phi_1}+e^{-i\phi_2})],\\
l_3&=i\frac{kB}{s_0}\frac{e^{-i\phi_1}+e^{-i\phi_2}}{-4-4\eta\bar{N}+4\eta\bar{N}\cos(\phi_1-\phi_2)}.
\end{aligned}
\end{equation}
To find the eigenbasis of the SLD, we diagonalize $\mathcal{L}_{\theta_1}$. Assuming $d_1=\frac{1}{\sqrt{2}}(a_1+e^{i\delta}a_2)$, $d_2=\frac{1}{\sqrt{2}}(a_1-e^{i\delta}a_2)$ and dropping the constant terms, we have
\begin{equation}
\mathcal{L}_{\theta_1}=(l_3e^{i\delta}+l_3^*e^{-i\delta})d_1^\dagger d_1-(l_3e^{i\delta}+l_3^*e^{-i\delta})d_2^\dagger d_2
+(l_3e^{i\delta}-l_3^*e^{-i\delta})d_1^\dagger d_2-(l_3e^{i\delta}-l_3^*e^{-i\delta})d_2^\dagger d_1.
\end{equation}
We can choose $l_3e^{i\delta}-l_3^*e^{-i\delta}=0$ or equivalently $\delta=\frac{1}{2}(\phi_1+\phi_2)-\frac{\pi}{2}$, which then means the SLD has the Fock basis of $d_1$, $d_2$ as its eigenbasis.  Thus the optimal POVM for estimating $\theta_1$ is $\{\ket{m,n}\bra{m,n}_d\}_{\{m,n\}}$, with $d_1^\dagger d_1\ket{m,n}_d=m\ket{m,n}_d$ and $d_2^\dagger d_2\ket{m,n}_d=n\ket{m,n}_d$. 

Notice the optimal POVM constructed above for the estimation of the centroid also depends on the centroid itself, so a different method would be used to measure the centroid. For example, just choosing the phase delay to be $\delta=0,\pi/2$ can be a method to find the centroid. Although this method is not optimal, unlike the separation, there is no fundamental limit to prevent us from improving the accuracy of estimating the centroid.

\section{Comparison with superresolution for resolving a weak thermal source}\label{weak}

We discuss superresolution for a weak thermal source in this section, which partially overlaps with the discussion in Ref. \cite{lupo2020quantum}. We show that the results in the main text can be reduced to the results for weak thermal sources in the weak source limit. Similar to Ref. \cite{tsang2011quantum}, we write down the received state from the source as 
\begin{equation}
\begin{aligned}
\rho&=(1-\epsilon)\ket{00}\bra{00}+\frac{\epsilon}{2}[\ket{01}\bra{01}+\ket{10}\bra{10}+g^*\ket{01}\bra{10}+g\ket{10}\bra{01}]+O(\epsilon^2)\\
&=(1-\epsilon)\rho_0+\epsilon\rho_1+O(\epsilon^2),
\end{aligned}
\end{equation}
where $g=\frac{1}{2}(e^{i\phi_1}+e^{i\phi_2})$, which encodes the information about the positions of the two sources. For a weak thermal source, $\epsilon\ll 1$ and thus the higher order terms can be ignored; then the quantum state and the measurement POVM are on a space spanned by $\ket{m,n}$ with $m+n\le 1$.

We then diagonalize the density matrix $\rho_1$ as 
\begin{equation}
\rho_1=D_1\ket{e_1}\bra{e_1}+D_2\ket{e_2}\bra{e_2},
\end{equation}
\begin{equation}
D_{1,2}=\frac{1}{2}\pm\frac{1}{4}e^{-i(\phi_1+\phi_2)/2}(e^{i\phi_1}+e^{i\phi_2}),
\end{equation}
\begin{equation}
\ket{e_{1,2}}=\pm\frac{1}{\sqrt{2}}e^{i(\phi_1+\phi_2)/2}\ket{01}+\frac{1}{\sqrt{2}}\ket{10}.
\end{equation}
The SLDs for both parameters $\theta_1$ and $\theta_2$ are then calculated as 
\begin{equation}
\begin{aligned}
\mathcal{L}_{\theta_1}&=-i\frac{kB}{s_0}\cos\frac{\phi_1-\phi_2}{2}\ket{e_1}\bra{e_2}+i\frac{kB}{s_0}\cos\frac{\phi_1-\phi_2}{2}\ket{e_2}\bra{e_1},
\end{aligned}
\end{equation}
\begin{equation}
\begin{aligned}
\mathcal{L}_{\theta_2}&=-\frac{kB}{4D_1s_0}\sin\frac{\phi_1-\phi_2}{2}\ket{e_1}\bra{e_1}+\frac{kB}{4D_2s_0}\sin\frac{\phi_1-\phi_2}{2}\ket{e_2}\bra{e_2}.
\end{aligned}
\end{equation}
We then calculate the QFI as 
\begin{equation}
F=\frac{k^2B^2}{s_0^2}\left[\begin{matrix}
\cos^2\frac{\phi_1-\phi_2}{2} & 0\\
0 & \frac{1}{4}
\end{matrix}\right],
\end{equation}
where $F_{ij}=\Tr(\mathcal{L}_{\theta_i}\mathcal{L}_{\theta_j}\rho)$. It is clear that when the separation vanishes, $\theta_2\rightarrow0$, and the QFI $F_{22}$ for estimating the separation $\theta_2$ also remains a constant. This shows we can avoid Rayleigh's limit for observation of weak thermal sources using interferometer arrays.  The POVM to achieve the superresolution predicted by the quantum Cramer-Rao bound can be found from the eigenbasis of $\mathcal{L}_{\theta_2}$, which is the projective measurement $\{\ket{e_1}\bra{e_1},\ket{e_2}\bra{e_2}\}$.

Note the implementation of the optimal POVM  $\{\ket{e_1}\bra{e_1},\ket{e_2}\bra{e_2}\}$ requires us to know information about $(\phi_1+\phi_2)/2$, which means we need to know the centroid of the two point sources. If the accuracy of estimating the centroid is not infinite, the sensitivity of estimating the separation is degraded; we now consider the dependence of the FI on misalignment of the centroid. Using the POVM $\{\ket{\widetilde{e}_1}\bra{\widetilde{e}_1},\ket{\widetilde{e}_2}\bra{\widetilde{e}_2}\}$, where $\ket{\widetilde{e}_{1,2}}=\pm\frac{1}{\sqrt{2}}e^{i(\phi_1+\phi_2)/2+\xi}\ket{01}+\frac{1}{\sqrt{2}}\ket{10}$ and $\xi$ quantifies the deviation of aligning the measurement due to the finite accuracy of knowing the centroid, the FI is degraded to be 
\begin{equation}
I_{22}=\frac{\cos^2\xi\sin^2\frac{\phi_1-\phi_2}{2}}{1-\cos^2\xi\cos^2\frac{\phi_1-\phi_2}{2}}\frac{k^2B^2}{4s_0^2}.
\end{equation}
If $\xi$ is nonvanishing, when the separation $\theta_2$ goes to zero and hence $\phi_1-\phi_2\rightarrow 0$, we find $I_{22}\rightarrow 0$. So just as the case for a single lens \cite{tsang2016quantum}, the superresolution in this limit relies on the assumption that we know the centroid perfectly and align the measurement device with perfect accuracy. However, the sensitivity of estimating the separation with a limited length of baseline can be achieved by improving the estimation of the centroid and accuracy of aligning the measurement device. There is no longer a fundamental reason, such as Rayleigh's limit, that prevents us from improving the sensitivity of estimating the separation.

 We now discuss the estimation of the centroid. The eigenbasis of $\mathcal{L}_{\theta_1}$ is $\ket{g_{1,2}}=(\pm e^{i(\phi_1+\phi_2+\pi)/2}\ket{01}+\ket{10})/\sqrt{2}$, which suggests the optimal measurement is the projection onto state $\ket{g_{1,2}}\bra{g_{1,2}}$. If we again consider the deviation $\xi$ in the measurement, i.e. $\ket{g_{1,2}}=(\pm e^{i(\phi_1+\phi_2+\pi)/2+i\xi}\ket{01}+\ket{10})/\sqrt{2}$, the FI of estimating the centroid is 
\begin{equation}
I_{11}=\frac{\cos^2\xi\cos^2\frac{\phi_1-\phi_2}{2}}{1-\sin^2\xi\cos^2\frac{\phi_1-\phi_2}{2}}\frac{k^2B^2}{s_0^2}=\frac{k^2B^2}{s_0^2}\left[1-\frac{1-\cos^2\frac{\phi_1-\phi_2}{2}}{1-\sin^2\xi\cos^2\frac{\phi_1-\phi_2}{2}}\right].
\end{equation}
When the deviation $\xi\neq 0$, the FI decreases, which degrades the sensitivity. But unlike the estimation of the separation, the FI of the centroid estimation is always a finite value even if the separation goes to zero ($\phi_1-\phi_2\rightarrow 0$). Hence it is possible to have the variance of estimating the centroid to be at least as small as  $s_0^2/(k^2B^2)$, which does not depend on the separation.  
 Since the phase delay used to estimate the centroid actually requires information about the centroid, we might want to use an adaptive method that allows the phase delay to gradually approach the optimal value. This method would enhance the sensitivity compared to using a fixed, non-optimal phase delay that is arbitrarily chosen. 

We now consider the off-diagonal elements of the FI. If we use the projective measurement onto state $\ket{h_{1,2}}=(\pm e^{i\delta} \ket{01}+\ket{10})/\sqrt{2}$, the off-diagonal elements of the FI are
\begin{equation}
I_{12}=I_{21}=\frac{k^2B^2}{8s_0^2}\frac{\sin(\phi_1-\phi_2)\sin(\phi_1+\phi_2-2\delta)}{1-\cos^2\frac{\phi_1-\phi_2}{2}\cos^2(\frac{\phi_1+\phi_2}{2}-\delta)}.
\end{equation}
We can see (only) in the case $\delta=(\phi_1+\phi_2)/2$ do we have $I_{12}=I_{21}=0$, which means there are no statistical correlations between the separation and the
centroid \cite{ragy2016compatibility}. But as pointed out above, to choose $\delta=(\phi_1+\phi_2)/2$, we need perfect knowledge of the centroid and any error in our knowledge of the centroid will still degrade the estimation of the separation.
For any other $\delta$, the error of estimating the separation or the centroid   deteriorates the precision of estimating the other parameter due to their statistical correlation. Since the calculation here can be regarded as a limiting case of the thermal source of arbitrary strength discussed in the main text, we might expect a similar result for the thermal source of arbitrary strength, namely that the estimation of the separation will be degraded by the error of estimating the centroid, and vice versa.

\section{Comparison with the conventional method}\label{compareconventional}

\begin{figure}[!htb]
\begin{center}
\includegraphics[width=0.5\columnwidth]{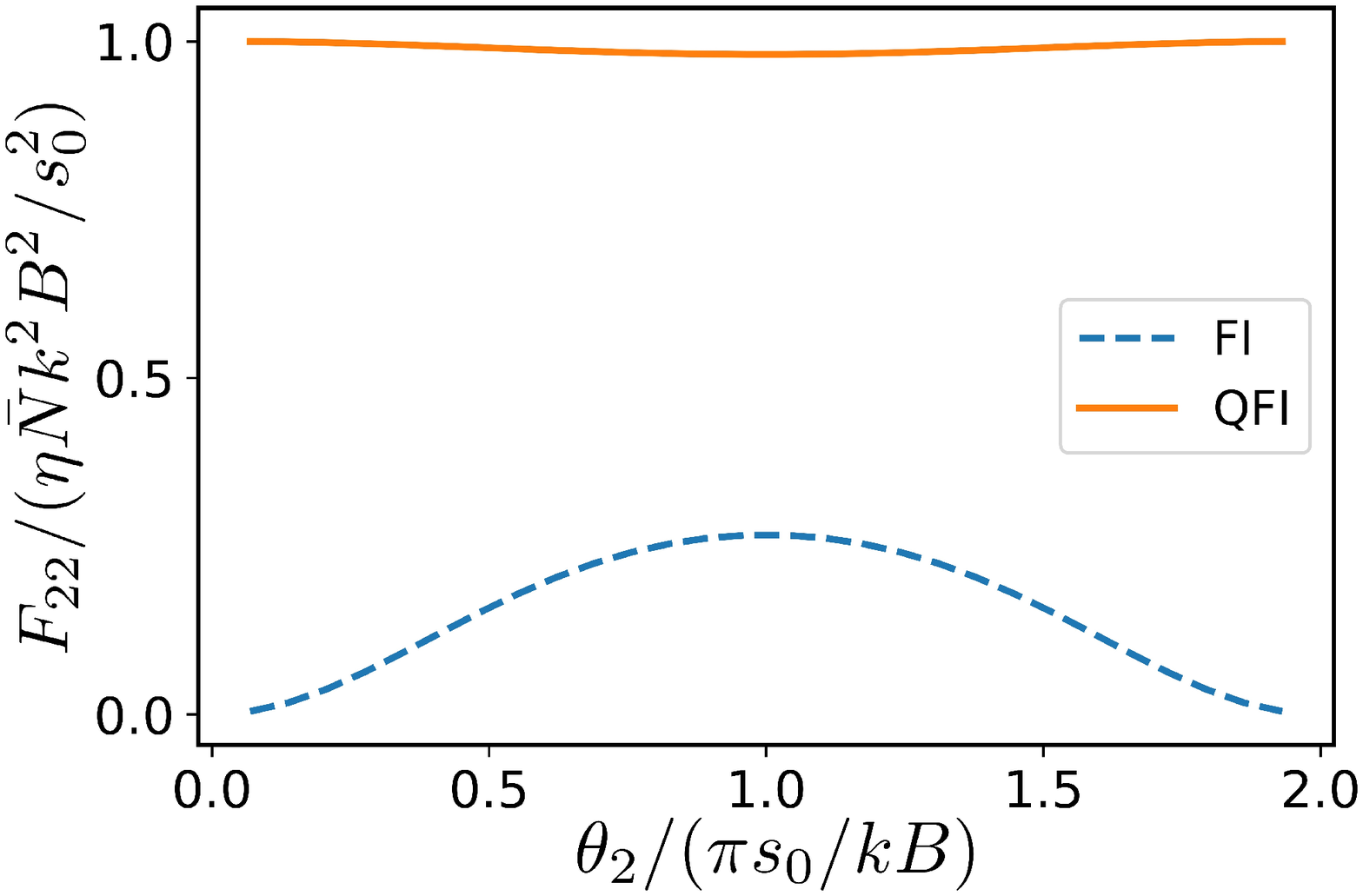}
\caption{Approximate Fisher information (blue dashed curve) and the quantum Fisher information (solid orange curve) as a function of separation $\theta_2$. Other parameters are chosen as $\bar{N}=0.01$ and $m,n\leq 3$. } 
\label{misalignment_compare}
\end{center}
\end{figure}

To compare the conventional method with our method, we assume the centroid is $\frac{1}{2}(\phi_1+\phi_2)=2\pi/3$ and choose the phase delay $\delta=0,\frac{\pi}{2}$, which is conventionally used to extract information about the coherence function. We calculate the FI for the POVM with phase delay $\delta=0,\frac{\pi}{2}$, corresponding to $c=2\pi/3,-\pi/3$. The FI and QFI are shown in Fig.~\ref{misalignment_compare} as a function of separation. It is clear that when the separation $\theta_2$ tends to zero, the FI vanishes, which implies the resolution limit, but the QFI remains a constant. This shows that a better POVM, such as the POVM we construct, can help avoid this limit. 

We can read from Fig.~\ref{misalignment_compare} that the conventional method requires the separation to be comparable to $s_0/kB$ to get reasonable sensitivity. This is consistent with the fact that the angular resolution of an interferometric array is approximately $\lambda/B$, where $\lambda$ is the wavelength of the received state. As a practical example, we consider the case where the observation is made with wavelength $\lambda=5$~mm and longest baseline $B=10$~km. Then the resolution of the conventional method is $\lambda/B=5\times 10^{-7}~ \mathrm{radians}\approx 0.1''$. For this case, when the angular separation of the two point sources is $\theta_2/s_0=0.05''$ and $\eta \bar{N}=0.01$, the Fisher information of our optimal measurement is larger than the conventional method by a factor of roughly 4. If we assume the mean square error of estimating the angular separation $\theta_2/s_0$ scales with the number of samples $n$ as $\Delta (\theta_2/s_0)^2\propto 1/n $,  this implies that our optimal measurement can shorten the observation time by a factor of 4 to achieve the same sensitivity. When the angular separation of the two point sources is $\theta_2/s_0=0.01''$, the Fisher information of our optimal measurement is larger than the conventional method by a factor of roughly 30, which shortens the observation time by a factor of 30. Finally, when the angular separation of the two point sources is $0.005''$, the Fisher information of our optimal measurement is larger than the conventional method by a factor of roughly 100, which shortens the observation time by a factor of 100.

\section{Multiple sources and detectors}\label{multiple}
Here we discuss extending our results to multiple sources and detectors. We assume the states emitted by the $s$th point source are thermal states $\rho^{th}$ with mean photon number $\bar{N_s}$, $s=1,2,\cdots, M$. The thermal states of modes $\{c_s\}$ corresponding to all point sources are described in Ref.  \cite{weedbrook2012gaussian} as
\begin{equation}
\begin{aligned}
\rho=\bigotimes_s\rho^{th}(\bar{N}_s)
=\prod_s\frac{1}{\pi\bar{N}_s}\int_{C^2}\prod_sd^2\alpha_s \exp \left(-\frac{\abs{\alpha_s}^2}{\bar{N}_s}\right)\bigotimes_s\ket{\alpha_s}\bra{\alpha_s}_c,
\end{aligned}
\end{equation}
where $\ket{\alpha_s}$  is the coherent state of $c_s$. We use the transformation from $c_s$ to the detector modes $a_j$, $j=1,2,\cdots, n$, 
\begin{equation}
\begin{aligned}
&c_s\rightarrow \sum_{j=1}^n\sqrt{\eta_{sj}}e^{i\phi_{sj}}a_j+\sqrt{1-\eta_s}v_s,\\
\end{aligned}
\end{equation}
where $\eta_s=\sum_j\eta_{sj}$ is the total loss from the source to the detector, $\eta_{sj}$ quantifies the loss from source $s$ to detector $j$, and $\phi_{sj}$ is the phase accumulated in propagation from source $s$ to detector $j$. In the far-field limit, $\phi_{sj}=k(u_jx_s+v_jy_s)/s_0$ for source $c_s$ with two-dimensional (2D) coordinate $(x_s,y_s)$ on the source plane and detector $a_j$ with 2D coordinate $(u_j,v_j)$ on the detection plane as derived in Ref.~\cite{lupo2020quantum}. The mean displacement $\lambda_\mu$ is still zero for all $\mu$. The covariance matrix of this state can be similarly derived as 
\begin{equation}
\Sigma=\left[\begin{matrix}
P_1 & Q_{12} & Q_{13} & \cdots & Q_{1n}\\
Q_{21} & P_2 & Q_{23} & \cdots & Q_{2n}\\
Q_{31} & Q_{32} & P_3 & \cdots & Q_{3n}\\
\cdots & \cdots & \cdots & \cdots & \cdots \\
Q_{n1} & Q_{n2} & Q_{n3} & \cdots & P_n\\
\end{matrix}\right],
\end{equation}
where $P_l$ and $Q_{lm}$ are $2\times 2$ matrices,
\begin{equation}
P_l=\left[\begin{matrix}
0 & \frac{1}{2}+\sum_s\eta_{sl}N_s\\
\frac{1}{2}+\sum_s\eta_{sl}N_s & 0\\
\end{matrix}\right],\quad Q_{lm}=\left[\begin{matrix}
0 & \sum_s\sqrt{\eta_{sl}\eta_{sm}}N_se^{i\phi_{sm}-i\phi_{sl}}\\
 \sum_s\sqrt{\eta_{sl}\eta_{sm}}N_se^{-i\phi_{sm}+i\phi_{sl}} & 0
\end{matrix}\right].
\end{equation}
The QFI and SLD can be found using Eq.~\ref{QFI_formula} and Eq.~\ref{SLD_formula}, but we are unable to simplify the equations and give general analytical results because the inverse of the covariance matrix is hard to determine. But in principle one can use the same calculation to find the QFI and SLD at least numerically for the estimation of any position information.



\twocolumngrid

\end{document}